\documentstyle[12pt]{article}
\input {peter1style}
\input {peter2math_macros}

\begin{document}

\begin{titlepage}
\nopagebreak
\begin{flushright}

        {\normalsize       OU-HET-178  \\
						   UTTG-14-93 \\
                           May~1993\\}

\end{flushright}

\vfill
\begin{center}
{\large \bf Universal and Nonperturbative Behavior  \\
  in the One-Plaquette Model of \\
 Two-Dimensional String Theory}

\vfill
       {\bf S.~Chaudhuri}

       Theory Group, RLM 5.208 \\
	   Department of Physics, \\
       University of Texas, \\
       Austin,  TX 78712  USA\\

\vfill
        {\bf H.~Itoyama   and  T.~Ooshita}

        Department of Physics,\\
        Faculty of Science, Osaka University,\\
        Toyonaka, Osaka, 560 Japan\\

\vfill
\begin{abstract}
The one-plaquette Hamiltonian of large N lattice gauge theory offers a
constructive model of a $1+1$-dimensional string theory with a stable ground
state.
  The free energy  is found to be equivalent
 to  the partition function of a string where the world sheet is discretized
 by even polygons with signature and the link factor is given by
  a non-Gaussian propagator.
 At large, but finite, N we derive the nonperturbative density of
states from the WKB wave function and the dispersion relations.
This is expressible as an infinite, but convergent, series with the inverse
of the hypergeometric function replacing the harmonic oscillator spectrum
of the $1+1$-dimensional string. In the scaling limit,
 the series is shown to be
finite, containing both the perturbative (asymptotic) expansion of the
inverted harmonic oscillator model, and a nonperturbative piece that
survives the scaling limit.

\end{abstract}
\end{center}
\vfill
\end{titlepage}

\section{Introduction}

Continuing efforts in string theory have shown that there are many
physical phenomena which may be described by strings but
in which  an understanding of nonperturbative effects are crucial.
A general formulation to handle these would require us to surpass
the conventional first quantized approach to strings.
However in the context of certain discretized toy models
it has been possible to resum string perturbation theory
provided the central charge $c$ of the matter system is
less than one \cite{BKDSGM}. Matrix models provide a powerful computational
tool to study these systems. Families of classical integrable differential
equations have been  found to govern the space of correlation functions.

The borderline case in which the central charge is equal to
one has also been studied in much detail \cite{c=1,Review},
and is of greater physical interest as a $1+1$-dimensional
  string theory than the $c< 1$ models.
Universality tells us that the characteristic critical behavior
can be extracted from a one-dimensional matrix model with
an inverted Gaussian potential. An expression for the density of states
up to arbitrarily high genus has in fact been obtained by this method
in the double scaling limit and is represented as a divergent infinite
series \cite{c=1}. This expansion is asymptotic as well as perturbative
and must be amputated to obtain a finite numerical value.
(The model is perturbatively stable since the instability due to tunneling
through the barrier is exponentially suppressed for large $N$.)
It has proven difficult to define a
 model \cite{Kaz,Moore}
 with a genuinely nonperturbative piece in
the density of states purely on the basis of an asymptotic expansion.
The point of view pursued in this paper is that this divergence arises
due to the absence of a stable ground state.

The one-plaquette Hamiltonian of large N lattice gauge theory offers a
genuinely nonperturbative and stable model of such a $1+1$-dimensional
string theory. Here the gauge group is taken to be $U(N)$.
One replaces the potential of the inverted harmonic oscillator
by a periodic potential which is bounded from below.
The model displays the third order large $N$ phase transition
of two-dimensional lattice QCD \cite{GrossWitten}. It defines the limiting
point of the unitary one-matrix models in zero dimensions \cite{umm}
\cite{neub} which, in a certain phase, display critical behavior in the
same universality class as the hermitian matrix models \cite{dem}.
Simple classical intuition tells us that the one-plaquette model shares
the same critical singularity as the hermitean matrix model with a gaussian
interaction. This well-known correspondence leads to a model of a string
theory with a natural nonperturbative regulator without introducing any
additional parameter \cite{Kaz}.

In this paper, we derive the nonperturbative expression for the
density of states for this model. This will be done first
at large, but finite, $N$ starting with the WKB wave function. The
matching of the WKB wave function over the periodic domain
 and subsequently the Hilbert
transform of the coefficient functions
are the two essential steps of our calculation.
Our formula is expressible as an infinite, but
convergent, series with the inverse of a hypergeometric function
replacing the harmonic oscillator spectrum.
In the scaling limit, we find that the series is still finite,
and contains the perturbative expansion coming from the inverted harmonic
oscillator. We are, therefore, able to identify a nonperturbative piece
that survives the scaling limit.

In section $2$, we review the definition of the Hamiltonian
one-plaquette lattice gauge theory.  We find that
 it  can be represented as a discretized model of a $1+1$-dimensional string
theory.  We review the reduction of the eigenvalue problem to the
solution of the one-body problem defined by Mathieu's equation.
We then calculate the properties of the ground state, obtaining
the planar singularity \cite{Wadia}
 and, in the double scaling limit, a recursive
evaluation of the perturbation series for the density of states function.

Section $3$ presents the main results of our work.
We first construct the semi-classical eigenfunction of
the one-body problem explicitly. The function is real, and periodic
with period $2\pi$, and possesses definite parity.
These properties are in accordance with the known results on the
Mathieu functions.  The construction provides the absolute values of
the coefficients of the wave function as a function of the one-particle
energy level $\mu$. These coefficients are the counterparts of the
reflection and absorption coefficients of the corresponding scattering
problem.  The Hilbert transform of the absolute value gives us
an expression for the density of states through the Bohr-Sommerfeld
quantization condition.

We then prove the finiteness of the density of states in the scaling limit.
We find that the interchange of the two limits leads to
the known perturbative expression. The perturbative series
is thus shown to be contained in our formula.
The interchange of the limits enables us to exhibit the
perturbative piece by blowing up the region of the planar singularity.
Finally in this section, we present the nonperturbative
expression of the scaled ground state energy as an implicit function of
the renormalized cosmological constant.
Section $4$ contains a brief summary of our conclusions.
The original model defines a Hamiltonian formulation of two-dimensional
$U(N)$ lattice gauge theory and may also serve as a toy model for exploring
some of the issues that arise in interpreting two dimensional QCD as
a theory of maps \cite{qcd2old} \cite{qcd2new}.

\section{The One-Plaquette  Model}
Before presenting our main results in the next section
we introduce the Hamiltonian one-plaquette model. We find a representation
of the model  as a $1+1$ dimensional string theory. We then review
the formalism for evaluating
the properties of the ground state and
obtain the planar singularity and the existence
of strong and weak coupling phases \cite{Wadia,JS}.
The vanishing of the mass gap at criticality was noted in \cite{neub},
which also contains a discussion of tunneling contributions.
Finally, we evaluate the density of states function in the double scaling
limit, recursively, as a perturbative expansion in the
string coupling following \cite{c=1}.
This can be done in parallel to what has been discussed in the
hermitean one-dimensional matrix model and our discussion below will
therefore be brief.

Let us first introduce the model as a large $N$ lattice gauge Hamiltonian
defined on a space-like plaquette.  The Hamiltonian  is given in terms
of the four oriented link variables $U_{1}, U_{2}, U_{3}$ and $U_{4}$ :
\beqn
\label{eq:1PKS}
      H = \left( g^{2}/2 \right)  {\displaystyle
 \sum_{\alpha, \ell}   E_{\ell \alpha}   E_{\ell \alpha}  }
             + g^{-2} \left( 2N -Tr U_{1} U_{2} U_{3} U_{4} -
  Tr U^{\dagger}_{4} U^{\dagger}_{3}
U^{\dagger}_{2} U^{\dagger}_{1}  \right)\;\;\;,
\eeqn
where  $ E_{\ell \alpha} = tr t^{\alpha} U_{\ell} \frac{d}{dU_{\ell}}  $.
Gauge invariance states that physical states must be invariant under
the local gauge transformations
$   U_{\ell} \rightarrow
 V_{\ell-1, \ell} U_{\ell} V_{\ell, \ell+1}^{\dagger}$:
\beqn
\label{eq:physt}
 \Psi \left (   V_{\ell-1, \ell} U_{\ell} V_{\ell, \ell+1}^{\dagger}   \right)
  = \Psi \left (    U_{\ell}    \right) \;\;\;
\eeqn
which implements the Gauss law constraints.
Acting on the gauge invariant subspace, we can reduce  the
form of the Hamiltonian to
\beqn
\label{eq:redham}
    H = -4\left( g^{2}/2 \right) \Delta + g^{-2} \left( 2N -Tr U -
  Tr U^{\dagger} \right)\;\;\;.
\eeqn
  Here  $U = U_{1} U_{2} U_{3} U_{4}$ and
 $\Delta$ is the Laplace operator acting on the unitary
 group. It can be decomposed in the angular momentum basis as
\beqn
\label{eq:decomposition}
  \Delta  =   \frac{1}{J} {\displaystyle \sum_{i=1}^{N}
  \frac{\partial^{2}}{\partial \theta_{i}^{2} }
 \left( J \right) +
   \frac{1}{4}   \sum_{i \neq j}^{N}  L_{ij} L_{ij}^{*}  cosec^{2}
   \frac{1}{2} \left( \theta_{i} - \theta_{j} \right)  }  \\
   J =  {\displaystyle \prod_{i < j} 2 \sin  \frac{1}{2} \left(
  \theta_{i} -\theta_{j} \right) \;\;,\;\;
    [ L_{ij},  L_{\ell m}] = -i \left(
 L_{im}  \delta_{\ell j} -  L_{\ell j}   \delta_{i m} \right) }  \;\;\;.
\eeqn

One can reinterpret this Hamiltonian as arising from the partition
function describing the one-dimensional quantum mechanics of an $N \times N$
unitary matrix
\beqn
\label{eq:Z}
  Z_{N} =  \int {\cal D}U \exp
 \left[ -  \frac{1}{8g^{2}} \left\{ \int dt tr  \left( \dot{U}^{\dagger}
 \dot{U}  \right)
    +8 tr \left( 2 - U - U^{\dagger} \right) \right\} \right] \;\;\;.
\eeqn
Here $  {\cal D}U $ denotes ${\displaystyle \prod_{t \in {\bf R}} dU(t)}$
which is the product of the $U(N)$ invariant Haar measure
over time $t$. The compactness of the group ensures that
the integration is well-defined and finite,
which is in contrast to the divergent integral seen in the hermitean
matrix case. Let us see how the free energy of this system
is related to a string partition function having a discretized
world sheet. Define the $N$ by $N$ hermitean matrix $M(t)$ by
$\ln U(t)  \equiv 2ig \sqrt{N} M(t)$. We find that the free energy
is expressible as
\beqn
\label{eq:lnZ}
  \ln Z_{N}   = \sum_{h=0}^{\infty} N^{2-2h} \sum_{  \{ \{ G^{*} \}\} }
 \frac{(sgn)}{ \mid G^{*} \mid}
   \left( 4\lambda \right)^{
 \displaystyle {\sum_{\ell=2}^{\infty}  } \left(\ell -1\right)
 V_{2\ell} }  \prod_{i=1}^{V}  \left(8 \int dt_{i} \right)
  \prod_{<ij>} P(t_{i} -t_{j})
\eeqn
where the second summation implies the summation over the
graphs  $  \{ \{ G^{*} \}\}  $
with a fixed number of handles $h$.  We denote by   $V_{2\ell}
 ~~ (  \ell = 2,3,4, \cdots )$
the number of vertices of type $M^{2\ell}$ in graph $G^{*}$
and  $V= {\displaystyle \sum_{\ell=2}^{\infty} } V_{2\ell} $.
The sign factor $ (sgn) = \left(- \right)^{ \displaystyle
  \sum_{\ell=2}^{\infty} \ell V_{2\ell} }$   specifies a rule
 that assigns a factor of $(-1)$
 to every $M^{4\ell +2}$ vertex in the graph $G^{*}$. The factor
$\mid G^{*} \mid$ is the order of the symmetry group of the graph
and  $\lambda \equiv g^{2}N$. The propagator $P(t - t^{\prime})$ is defined by
\beqn
  <<  M_{ij} (t) M_{k\ell}(t^{\prime} ) >>
 \equiv \frac{1}{N} \delta_{i\ell}
 \delta_{jk}  P ( t - t^{\prime} )  \;\;\;,
\eeqn
where the averaging $<<...>>$ is done with respect to
\beqn
\label{eq:Uint}
 {\cal D}U \exp
 \left[ -  N \left\{ \int dt \frac{1}{8 \lambda} tr \dot{U}^{\dagger}
 \dot{U}
    + 4 tr M^{2}  \right\} \right] \;\;\;.
\eeqn
Note that the integrals in eq.~(\ref{eq:Uint}) are not of Gaussian form.
The propagator connects the two vertices $t$ and $t^{\prime}$
and $  {\displaystyle \prod_{<ij>}  }$ means  taking
products over all links in graph $G^{*}$.

Eq.~(\ref{eq:lnZ}) is interpreted as a partition function
of a discretized string with $8t_{i}$ being a classical embedding
of the string surface into the real line. The discretization of the surface
is done by the polygons with even numbers of vertices and is represented
by a graph $G$ which is dual to $G^{*}$. The area of a $2 \ell$-polygon
is taken to be $\ell -1$ in these units. The factor
$ \left( 4\lambda \right)^{\displaystyle
  \sum_{\ell=2}^{\infty} \left(\ell -1\right)
 V_{2\ell} }$ is written as
\beqn
     \exp \left[ \left(  \ln 4\lambda_{c} +
 a^{2} \tilde{x} \right)   \left( \sum_{\ell=2}^{\infty} \left(
  \ell-1\right) V_{2\ell} \right) \right] \;\;\;,
\eeqn
where $\lambda /\lambda_{c}  \equiv 1+x  \equiv 1+ a^{2} \tilde{x}$,
and $\lambda_{c}$ is the critical value of $\lambda$.
The combination $ {\displaystyle \sum_{\ell=2}^{\infty}  } \left(
  \ell-1\right) V_{2\ell} a^{2}$ is
regarded as a definition of the physical area of the surface
in the continuum limit $ a^{2} = 1/N \rightarrow 0$, and
$ \tilde{x}$ is the renormalized cosmological constant.
The presence of minus signs is closely related to the fact
that the right hand side of eq.~(\ref{eq:lnZ}) is convergent.
A minus sign is assigned to every $(4 \ell +2)$-polygon in the graph $G$.

In what follows, we will be interested in the ground state properties of
the Hamiltonian (eq.~(\ref{eq:redham})).
It is sufficient to restrict our attention to the eigenvalue problem
\beqn
   H \Psi = E \Psi
\eeqn
where we restrict ourselves to the singlet subspace, $L_{ij}  \Psi =0 $.
Namely,
\beqn
\label{eq:evp}
 -2g^{2} \frac{1}{J} {\displaystyle \sum_{i=1}^{N} \frac{\partial^{2}}
 { \partial \theta_{i}^{2}} \left( J \Psi \right ) + \frac{2N}{g^{2}}
\left( 1- \frac{1}{N} \sum_{i=1}^{N} \cos \theta_{i} \right ) \Psi
  = E \Psi  }  \;\;\;.
\eeqn
Eq.~(\ref{eq:evp}) is separable into a one-body problem.
Introducing  $  E/ N^{2} = \frac{1}{N}
{\displaystyle   \sum_{i=1}^{N} \epsilon_{i} }$ and
\beq
   \Psi \left( \theta_{1} \cdots \theta_{N} \right)
    =  \det  \left( \psi_{i} \left( \theta_{j} \right) \right) /
  J\left(  \theta_{1} \cdots \theta_{N} \right)\;\;\;,
\eeq
we reduce the problem to
\beqn
 \hat{h}_{0} \psi  \left( \theta \right) &=& \epsilon
 \psi
 \left( \theta \right). \;\;  \nonumber \\
  \hat{h}_{0} &\equiv&  -  \frac{2\lambda}{N^{2}}
 \frac{d^{2}}{d \theta^{2}}    + \frac{2}{\lambda} \left( 1-\cos
 \theta \right)   \;\;\;,
\eeqn
where the one-particle wave function is subject to periodic boundary
conditions, $\psi  \left( \theta
 + 2 \pi \right)  = \psi  \left( \theta \right).$

For our purpose, it is convenient to shift variables,
  $ \chi = \pi - \theta$, and define the shifted  one-particle energy,
 $ \mu = \epsilon - \frac{4}{\lambda} $. Then
\beqn
  \hat{h}   \psi  \left( \chi ; \mu \right) &=& \mu
 \psi  \left( \chi ; \mu \right)\;\;, \;\;\;\;
     \psi  \left( \chi  + 2 \pi \mu \right ) =
    \psi  \left( \chi ; \mu\right) \;\;\;,
 \nonumber \\
 \hat{h} &\equiv&   - \frac{2\lambda}{N^{2}} \frac{d^{2}}{d \chi^{2}}
     + \frac{2}{\lambda} \left( \cos
 \chi -1 \right)   \label{eq:mathieu} \;\;\;.
\eeqn
Eq.~(\ref{eq:mathieu}) is known as Mathieu's equation.

The density of one-particle states is defined by
\beqn
\label{eq:rho}
  \rho_{N} \left( \mu ; \lambda \right) \equiv
   \frac{1}{N} \frac{\partial n}{ \partial \mu} \equiv
   \frac{1}{N} {\displaystyle \sum_{i=1}^{ \infty} }
 \delta \left( \mu - \mu_{i} \right)
  =  \frac{1}{N} tr  \delta \left( \mu - \hat{h} \right)
  \;\;\;,
\eeqn
where $ n \equiv   \frac{1}{N}  \int^{\mu} d \mu^{\prime}
  tr  \delta \left( \mu^{\prime} - \hat{h} \right)$
is the number of energy levels up to $\mu$.
Since $n$ counts the number of zeroes of the $n^{th}$ eigenfunction,
eq.~(\ref{eq:rho}) is also expressible in terms of the function
$ w(\chi;\mu, \lambda)=- {{\lambda}\over{N}} {{\psi'(\chi ; \mu)}\over
{\psi(\chi ; \mu)}}$ as
\beqn
\label{eq:dense}
\rho(\mu ; \lambda) =  - \frac{1}{\lambda} \int_{-\pi}^{+\pi}
 {{d \chi}\over{2 \pi i}} {{\partial w(\chi;\mu , \lambda )}
 \over{\partial \mu}}  \;\;\;.
\eeqn
The latter expression can be used to generate a $\frac{1}{N}$ expansion
by solving for $w(\chi;\mu , \lambda )$ recursively in the Mathieu equation.

To see this, define $u=  \sin {{\chi}\over{2}} $. Then $w(u;\mu , \lambda )$
satisfies the first order Riccati equation
\beqn
\label{eq:riccati}
{{\lambda}\over{N}} \left [ (1-u^2)w'(u) - u w(u) \right ] =
(1-u^2)w^2 + 8 ( u^2 + {{\mu \lambda}\over{4}})
\eeqn
In the planar limit, setting the right hand side of the equation
 equal to zero and
substituting the resulting expression for $w$ in eq.~(\ref{eq:dense})
one obtains the integral representation of an elliptic function
\beqn
   \lim_{N \rightarrow \infty}
 \rho \left( \mu ;  \lambda \right)  =
    \frac{1}{ \sqrt{2} \pi}   \int_{0}^{\pi/2}  \frac{ d \chi^{\prime}}{2\pi}
   \frac{ \Theta \left(  \frac{\mu \lambda}{4} + \cos^{2} \chi^{\prime}
 \right)  }
   {  \sqrt{ \left(  \frac{\mu \lambda}{4} + \cos^{2} \chi^{\prime}
 \right) }  }
  \;\;\;.
\eeqn
Depending upon whether $\mu = \epsilon - \frac{4}{\lambda}$
is positive or negative, we obtain a different support for the integrand.
Defining the couplings,
$k_{s}^{2}  \equiv \left( \epsilon \lambda /4  \right)^{-1}$,
and $k_{w}^{2}  \equiv \left( \epsilon \lambda /4  \right)$ respectively,
and approaching the critical value from below in either case, we find
\beqn
\label{eq:planarrho}
      \frac{1}{ \sqrt{2} \pi}  k_{s}  K \left( k_{s} \right)
  ~~~ \Rightarrow_{k_{s} \rightarrow 1}~~~    \frac{1}{2 \sqrt{2} \pi}
   \log   \left( \frac{1}{ 1-k_{s}^{2} } \right)  \;\;\;, \\
    \frac{1}{ \sqrt{2} \pi}  K\left( k_{w} \right)
 ~~~ \Rightarrow_{k_{w} \rightarrow 1 }~~~    \frac{1}{2 \sqrt{2} \pi}
   \log   \left( \frac{1}{ 1-k_{w}^{2} } \right)  \;\;\;.
\eeqn
In the planar limit, the normalization condition
\beqn
\label{eq:norm}
 1 = \int_{-4/\lambda}^\mu d \mu^{\prime}
 \rho_{N}(\mu^{\prime};\lambda)  \equiv  {\cal N}_{N} \;\;\;
\eeqn
 reduces to
\beqn
\label{eq:normpl}
\frac{\lambda}{\lambda_{c}} &=&     \frac{1}{k_{s}}
  \int^{\pi/2}_{0} d\chi \sqrt{ 1+  k_{s}^{2}  \cos^{2} \chi } \;\;, \;
  {\rm for}~ k_{s}^{2} < 1 \;\;    \\
\frac{\lambda}{\lambda_{c}} &=&~
  \int^{\sin^{-1} k_{w}}_{0} d\chi \sqrt{   k_{w}^{2} -  \sin^{2} \chi } \;\;,
   \; {\rm for}~ k_{w}^{2} < 1 \;\;\;,
\eeqn
where  $ \lambda_{c} =  \frac{4 \sqrt{2}}{\pi}$ is the critical value.
In the weak coupling regime the wave function is localized
while in the strong coupling regime the wave function is spread
evenly over the domain. The difference in the support of the integrands
of eqs.~(\ref{eq:planarrho}),(\ref{eq:normpl}) is a consequence of this fact.
Introducing the scaling variable,
   $ x  \equiv    \frac{\lambda - \lambda_{c}}{\lambda_{c}}$, we obtain
\beqn
\label{eq:inversion}
  1-k_{s}^{2}  \sim     \frac{4x}{ \log \mid 1/4x \mid}\;\;\;\;
  1-k_{w}^{2}  \sim   \frac{ -4x}{ \log \mid 1/4x \mid}\;\;\;\;.
\eeqn

The $\frac{1}{N}$ corrections to this result in the usual double
 scaling limit,
$N \rightarrow \infty$, $\eta={{\mu \lambda}\over{4}} \rightarrow 0 $, with
$\alpha = -{{\lambda}\over{2\sqrt 2 N \eta}} $ held fixed, are identical to
those in the inverted harmonic oscillator model. To see this, define
$v(u) \equiv \sqrt{1-u^2} ~ w(u)$ and
$- \eta t^{2} \equiv 1- u^{2}$ in eq.~(\ref{eq:riccati}) which leads
to the equation
\beqn
\label{eq:dsc}
- \alpha \sqrt{1+ \eta t^2} {{d v}\over{d t}} + v^2 = 1+ t^2
\eeqn
which coincides with that derived in  the last reference of
 \cite{c=1} at criticality
$\eta = 0$. The solution can be obtained by recursively solving for
the coefficients in a trial solution of the form
 $ v(\alpha)=  {\displaystyle \sum_k  } c_k \alpha^k $
to eq.~(\ref{eq:dsc}).
Not surprisingly, once we approach criticality, we recover the
 well-known results of the one-dimensional
 hermitean matrix model  \cite{c=1,Moore}.

In the next section, we will solve first for the density of states at
finite $N$ in closed form
  and then observe that there is, in fact, a non-trivial scaling
limit for this solution in which the nonperturbative contribution is
 made explicit.

\section{Nonperturbative Formula for the Density of States}

Now we proceed to give a nonperturbative formula for the density of states
from the eigenfunctions of the Hamiltonian given by eq.~(\ref{eq:mathieu})
in the WKB approximation.
The factor $1/N$ plays the role of the Planck constant here.
As our discussion goes through several steps, we will divide the analysis
into subsections.

\subsection{Construction of semi-classical eigenfunctions}

We denote by   $ \psi^{\left(\pm \right)} \left( \chi; \mu \right) $
 a pair of wave functions which are complex conjugate to  each other
and whose semi-classical form is given by
\beqn
   \psi_{WKB}^{\left(\pm \right)} \left( \chi ; \mu \right) &=&
  \tilde{p}^{-1/2}
    \exp \left( \pm i N
  \int_{\pi}^{\chi} \tilde{p} \left( \chi^{\prime} \right) d
 \chi^{\prime}  \right) \;\;\;  \nonumber \\
   \tilde{p} \left( \chi ; \mu \right)
 &=& \sqrt{ \frac{1}{2\lambda}} \sqrt {\frac{4}{\lambda}
  \sin^{2} \frac{\chi}{2} + \mu} \;\;\;.
\eeqn
 The semi-classical approximation
 is valid in the region satisfying
\beqn
\label{eq:deBroglie}
    \frac{d}{d\chi} \left( \frac{1}{N \tilde{p}} \right) << 1 \;\;\;.
\eeqn
  Here  $\frac{1}{N \tilde{p} }$
 is  the de Broglie wavelength.  Eq.~(\ref{eq:deBroglie}) reads
\beqn
 \frac{1}{4\lambda} \mid \frac{2}{\lambda} \sin \chi \mid
   <<  N \tilde{p}^{3} \left(\chi \right) \;\;\;.
\eeqn
 Except at the turning point $ \tilde{p} \left( \chi \right) =0$,
 eq.~(\ref{eq:deBroglie})
 can be satisfied for arbitrary $\chi$ by letting $N$
 get sufficiently large.
(A proposal for an improved WKB wavefunction whose region of validity
extends into the vicinity of the turning points of the potential was
given in \cite{neub}.)

 Expanding  $\tilde{p} \left( \chi ;\mu \right)$
 as a power series in $\mu \lambda/4$,
 and  integrating term by term, we find
\beqn
\label{eq:pdchi}
     \int_{\pi}^{\chi} d \chi^{\prime}
 \tilde{p} \left( \chi^{\prime} \right)
  =    R \left( \chi/2 \right) - \frac{ \sqrt{2}}{ \lambda} g \left(
 -\mu \lambda /4 \right) \log \tan \left( \chi /4 \right)  \;\;\;.
\eeqn
  Here
\beqn
\label{eq:R}
     R \left( \chi/2 \right) =  - \frac{2 \sqrt2}{ \lambda} \cos  \left(
 \chi /2 \right) +   \frac{ \sqrt2}{ \lambda} {\displaystyle
 \sum_{p=1}^{\infty}  } \frac{  \left( -\mu \lambda/4 \right)^{p}}{ p!}
 \frac{  \left( \left(2p-3\right)!! \right)^{2} }
{2^{2(p-1)} \left(p-1 \right)! }     \;\;\;  \nonumber \\
  \times {\displaystyle \sum_{r=0}^{p-2} }
 \frac{ \left( 2(p-1) -2r-2 \right) !!}
{   \left( 2(p-1) -2r-1 \right) !!}
 \cos \left( \chi /2 \right)
 \left( \sin \left( \chi /2 \right) \right)^{-2(p-1)+ 2r}
 \;\;\;,
\eeqn
 and
\beqn
\label{eq:g}
    g\left( x\right) \equiv  {\displaystyle \sum_{p=1}^{\infty} }
  \frac{x^{p}}{p!}
 \frac{ \left( \left(2p-3 \right)!! \right)^{2} }
{2^{2(p-1)} \left(p-1\right)!} \;\;\;.
\eeqn
   Eqs.~(\ref{eq:R}),(\ref{eq:g})
 are obtained by  integrations of the inverse powers of
  trigonometric functions.  We find that
 the infinite series $g\left(x\right)$, which
 is convergent at $\mid x \mid < 1$, is simply related to
 the hypergeometric  function
  $F( \alpha, \beta  ; \gamma  \mid x)$ by
\beqn
\label{eq:ghyper}
 g(x) = x  F( \alpha = 1/2, \beta = 1/2 ; \gamma =2 \mid x) \;\;\;.
\eeqn
This defines $g(x)$ beyond its  radius of convergence by  the integral
 representation of   $F( \alpha, \beta  ; \gamma  \mid x)$.
 The function $R$ satisfies
\beqn
  R\left(-\chi/2 \right) =  R\left(\chi/2 \right)
   \;,\;\;    R\left( \left( \chi + 2\pi \right)  /2 \right)
    = -  R\left(\chi/2 \right) \;\;\;.
\eeqn
 The multivaluedness under $\chi \rightarrow -\chi$ is solely due to
 the second term in eq.~(\ref{eq:pdchi}) and the  separation of the
  WKB phase factor  in eq.~(\ref{eq:pdchi}) is unambiguous.

It is interesting to compare eq.~(\ref{eq:pdchi}) with the corresponding
 expression in the case  of the inverted harmonic oscillator:
\beqn
  \int^{x} dx^{\prime} \tilde{p} (x^{\prime}) =
 x^{2} /2 + \mu \log x
\eeqn
  obtained in the appendix ( eq.~(\ref{eq:invwkb}) ).
 There the integration of $1/x$ is the only term
 which yields the logarithm.   In contrast,
  any inverse odd power of the  $\sin$ contains a logarithm.
 This  is  why we  obtain the infinite series
 of  eq.~(\ref{eq:g}).

 This difference can be understood in physical terms.
  As we will see later in this section,  the term   with
 the multivaluedness  in the exponential
 provides  matching  conditions on the WKB wave function   which then
 determine the density of states.
 Since the inverted harmonic oscillator is a bottomless potential,
no qualitative change in energy levels
 is expected to occur as one lowers  the
 energy $\mu$ from the top of the potential.
 The term in the WKB phase factor which multiplies $\log x$
 is simply $ \mu $  --- an entire function and no nonanalyticity
 with $ \mu $  has in fact been  produced.
In the case of the one-plaquette model,
 the potential is bounded from below and the energy levels are discrete
 without putting infinite walls. As one lowers the energy  $\mu$ from
the top of the potential,
 he starts seeing the bottom.
The expression  for the density of states should reflect this, which
 is accomplished by the infinite series $g \left(x\right)$.
 The convergence radius $ \mid x \mid =1$ corresponds to  the bottom of
 the potential.

As the potential is real, the eigenfunctions can be chosen to be real:
\beqn
\label{eq:psi1}
 \psi^{\left(R \right)}\left( \chi ;  \mu \right)
   &=& B \psi^{ \left(+ \right)}\left( \chi ;  \mu  \right)
     + B^{*} \psi^{\left(- \right)}
    \left( \chi ;  \mu \right) \;  ,\;\;\;
 0 < \chi < \pi \;, \;\;\;    \\
 \label{eq:psi2}
 \psi^{\left(L \right)}\left( \chi ;  \mu \right)
   &=& A \psi^{ \left(+ \right)}
\left( \mid \chi \mid ;  \mu \right)
     + A^{*} \psi^{  \left(- \right)}
    \left( \mid \chi \mid ;
  \mu \right) \;  ,\;\;\; 0 <  - \chi < \pi \;.  \;\;
\eeqn
In the region satisfying  eq.~(\ref{eq:deBroglie}),
 the wave functions
   $  \psi^{\left(R \right)}\left( \chi ;  \tilde{\mu} \right)$
 and
 $ \psi^{\left(L \right)}\left( \chi ;  \tilde{\mu} \right)$
 are  given  by
\beqn
\label{eq:psiWKB}
  \psi^{\left(R \right)}_{WKB} \left( \chi ;  \mu \right)
   &=&
    \frac{2\mid B \mid}{\sqrt{ \tilde{p}} }
  \cos \left( \arg B +  N \int^{\chi}_{\pi} \tilde{p} \left(\chi^{\prime}
\right) d \chi^{\prime}  \right)  \;\;\;, \\
  \psi^{\left(L \right)}_{WKB} \left( \chi ;  \mu \right)
   &=&
    \frac{2\mid A \mid}{\sqrt{ \tilde{p}} }
  \cos \left( \arg A +  N \int^{ \mid \chi \mid }_{\pi}
 \tilde{p} \left(\chi^{\prime}
\right) d \chi^{\prime}  \right)  \;\;\;
\eeqn
  respectively.

 The matching of the  wave functions,
$\psi^{\left(R \right)}\left( \chi ;  \mu \right)$ $\left( {\rm for}~
  0 < \chi <\pi
  \right) $,
 and the function
 $ \psi^{\left(L \right)}\left( \chi ;  \mu \right)$ $ \left(
   {\rm for}~  0 < - \chi < \pi \right)$  along the
 negative real axis
 is done in the complex $\chi$ plane. Using the fact that
 $\psi^{\left(+ \right)} \left( \psi^{\left(- \right)} \right)$ dominates
 in the  second (third) quadrant, we find
\beqn
\label{eq:matching}
  B \psi^{\left(+ \right)} \left( \chi = r e^{i \alpha} ; \mu
 \right) \mid_{\alpha  \rightarrow \pi}
  &=&   A \psi^{\left(+ \right)} \left( \mid \chi \mid = r
 ; \mu \right)  \;\;, \;\; \nonumber \\
  B^{*} \psi^{\left(- \right)} \left( \chi = r e^{-i \alpha} ; \mu
 \right) \mid_{\alpha \rightarrow \pi}
  &=&   A^{*} \psi^{\left(- \right)} \left( \mid \chi \mid =  r
 ; \mu \right)  \;\;.\;\;
\eeqn
 Both of eq.~(\ref{eq:matching}) provide the relation
\beqn
\label{eq:ABpi}
  A =  e^{-\pi  N G(\mu  ; \lambda) } B \;\;\;,
\eeqn
 where
\beqn
\label{eq:mutilde}
  G(\mu ; \lambda ) \equiv
 - \left( \frac{ \sqrt{2}}{\lambda} \right) g \left(
 -\mu \lambda /4 \right) \;\;\;.
\eeqn
We determine the norm of the coefficients
 $A$ and $B$ through the normalization condition:
\beqn
 1=  \int_{-\pi}^{0} d \chi
  \left( \psi^{\left(L\right)} \left( \chi ; \mu \right) \right)^{2}
  + \int_{0}^{\pi} d \chi
  \left( \psi^{\left(R \right)} \left( \chi ; \mu \right) \right)^{2} \;\;\;.
\eeqn
  We find
\beqn
\label{eq:ABabs}
     \mid A \mid^{2} &\approx&  \frac{1} {4}  \frac{e^{-\pi N G  }}
{ \cosh \pi  N G   }
    \left( \int^{\pi} d \chi  \frac{1}{\tilde{p} \left( \chi; \mu  \right) }
 \right)^{-1}  \;\;\;, \nonumber \\
     \mid B \mid^{2} &\approx&  \frac{1} {4}
   \frac{e^{\pi  N G  }}{ \cosh \pi  N G }
    \left( \int^{\pi} d \chi  \frac{1}{\tilde{p} \left( \chi; \mu \right) }
 \right)^{-1}  \;\;\;.
\eeqn

 Thus far, we have constructed  a   real and  correctly
 normalized WKB wave function.
  Let us further make it  either even or odd under
 $\chi \rightarrow -\chi$.
  Exact properties  of the Mathieu functions
   tell us that there
exist  a countably infinite number of even periodic eigenfunctions and
   another series of a countably infinite number of odd
 periodic eigenfunctions which are not degenerate with the even ones.
 Let $\chi_{r} =- \chi$. We add and subtract the ``parity conjugate''
 wave function to the original forms (eqs.~(\ref{eq:psi1}),(\ref{eq:psi2})),
 so that the resulting wave functions are even and odd respectively:
\beqn
\label{eq:Psi}
 \Psi^{\left(R \right)}_{\pm} \left( \chi ;  \mu \right)
   &=& B \psi^{ \left(+ \right)}\left( \chi ;  \mu  \right)
     + B^{*} \psi^{\left(- \right)}
    \left( \chi ;  \mu \right) \;  \;\;\;  \nonumber \\
     &\pm&  \left(  A \psi^{ \left(+ \right)}
\left( \mid \chi_{r} \mid ;  \mu \right)
     +  A^{*} \psi^{  \left(- \right)}
    \left( \mid \chi_{r} \mid ;
  \mu \right)  \right)  \, ,\;\;\; 0 <  \chi < \pi \;.\\
\Psi^{\left(L \right)}_{\pm} \left( \chi ;  \mu \right)
   &=& A \psi^{ \left(+ \right)}
\left( \mid \chi \mid ;  \mu \right)
     + A^{*} \psi^{  \left(- \right)}
    \left( \mid \chi \mid ;
  \mu \right) \;   \nonumber \;\;  \\
    &\pm&  \left(  B \psi^{ \left(+ \right)}\left( \chi_{r} ;  \mu  \right)
       +  B^{*} \psi^{\left(- \right)}
    \left( \chi_{r} ;  \mu \right) \right) \;   ,\;\;\; 0 < - \chi < \pi \;.
\eeqn
  Here $\pm$ denotes the evenness and the oddness
 of the wave function respectively.
  Similarly,
\beqn
\label{eq:PsiWKB}
  \Psi^{\left(R \right)}_{\pm~WKB} \left( \chi ;  \mu \right)
   &=&
    \frac{2\mid B \mid}{\sqrt{ \tilde{p}} }
  \cos \left( \arg B +  N \int^{\chi}_{\pi} \tilde{p} \left(\chi^{\prime}
\right) d \chi^{\prime}  \right)  \;\;\;   \nonumber \\
  &\pm&    \frac{2\mid A \mid}{\sqrt{ \tilde{p}} }
  \cos \left( \arg A +  N \int^{ \mid \chi_{r} \mid }_{\pi}
 \tilde{p} \left(\chi^{\prime}_{r}
\right) d \chi^{\prime}_{r}  \right)  \;\;\;,    \\
 \Psi^{\left(L \right)}_{\pm~WKB} \left( \chi ;  \mu \right)
   &=&
    \frac{2\mid A \mid}{\sqrt{ \tilde{p}} }
  \cos \left( \arg A +  N \int^{ \mid \chi \mid }_{\pi}
 \tilde{p} \left(\chi^{\prime}
\right) d \chi^{\prime}  \right)  \;\;\; \nonumber \\
  &\pm&
    \frac{2\mid B \mid}{\sqrt{ \tilde{p}} }
  \cos \left( \arg B +  N \int^{\chi_{r}}_{\pi} \tilde{p}
 \left(\chi_{r}^{\prime}
\right) d \chi_{r}^{\prime}  \right)  \;\;\;.
\eeqn
  We have deliberately used  $\chi_{r}$ to make explicit that  the matching
 is done at the complex $ \chi_{r}$ plane  for the part we have
 added or subtracted.
  We now define the extension of the
 wave function  to the entire complex plane using periodicity.
 In the regime $ \pi < \chi < 2\pi$, explicitly
\beqn
\label{eq:L1}
\Psi_{\pm}^{\left(L_{1} \right)}\left( \chi ;  \mu \right)
   &=& A \psi^{ \left(+ \right)}
\left( \mid \chi - 2\pi \mid ;  \mu \right)
     + A^{*} \psi^{  \left(- \right)}
    \left( \mid \chi -2\pi \mid ;
  \mu \right) \;   \nonumber \;\;  \\
    &\pm&  \left(  B \psi^{ \left(+ \right)}\left( \chi_{r} + 2\pi
 ;  \mu  \right)
        +  B^{*} \psi^{\left(- \right)}
    \left( \chi_{r} + 2\pi ;  \mu \right) \right) \; \;,
\eeqn
  so that
\beqn
 \Psi_{\pm}^{\left(L_{1} \right)}
 \left( \chi ;  \mu \right)
  ~=~ \Psi_{\pm}^{\left(L \right)}\left( \chi- 2\pi ;  \mu \right) \;\;\;,
      \;\; \pi < \chi < 2\pi  \;\;.
\eeqn
   In general,
  we reinforce the periodic boundary condition by
\beqn
\Psi_{\pm}^{\left(L_{m} \right)}\left( \chi ;  \mu \right)  &=&
\Psi_{\pm}^{\left(L \right)}
 \left( \chi - 2m \pi ;  \mu \right) \;,\;\;
 (2m-1) \pi < \chi < 2m \pi \;\;\;,  \\
\Psi_{\pm}^{\left(R_{m} \right)}\left( \chi ;  \mu \right)  &=&
\Psi_{\pm}^{\left(R \right)}
 \left( \chi -2m \pi ;  \mu \right) \;,\;\;
 2m \pi < \chi < (2m+1) \pi \;\;\;.
\eeqn

 Now consider the regime $0 <  \chi < 2 \pi$.  We have the two wave functions
$\Psi^{\left(R \right)}\left( \chi ;  \mu \right)$ and
 $\Psi^{\left(L_{1} \right)}\left( \chi ;  \mu \right)$
 overlapping with each other.
  The heart of the Bohr-Sommerfeld quantization is
 that they must coincide:
\beqn
\label{eq:BS}
\Psi^{\left(R \right)}_{\pm} \left( \chi ;  \mu \right)  =
\Psi^{\left(L_{1} \right)}_{\pm} \left( \chi ;  \mu \right) \;\;\;.
\eeqn
  Using   $   \psi^{ \left(  \pm \right)}
\left( \mid \chi - 2\pi \mid ;  \mu \right)
       =   \psi^{ \left( \mp \right)}
\left( \chi  ;  \mu \right) $,
 we find
\beqn
  A \pm B = \pm \left( A^{*}  \pm B^{*} \right) \;\;\;,
\eeqn
   which provides the quantization condition
\beqn
  \arg \left(  A + B\right)  = n_{e} \pi \;\;, \;\;\;
   \arg \left(  A - B\right)  =     \left(  n_{o} +
 1/2 \right) \pi \;\;\;,\;\; n_{e}, n_{o}   \in {\cal Z} \;\;\;.
\eeqn
  We also note from eq.~(\ref{eq:ABpi}) that
\beqn
    \arg \left(  A \pm B\right)  =   \arg A
 = \arg B = \frac{1}{2} \left( \arg A + \arg B \right)\;\;,
 \;\; {\rm mod}~ {\cal Z} \;\;\;.
\eeqn
 in the WKB approximation.

\subsection{ The  density of states at finite $N$}

  The density of states  is given by
\beqn
\label{eq:dsarg}
  \rho_{N} ( \mu )
 = \frac{1}{2 \pi N} \frac{\partial}{\partial \mu}
  \left( \arg A + \arg B \right) \;\;\;.
\eeqn
To obtain the expression for the density of states from
 eq.~(\ref{eq:ABabs}), we note that
 the real and the imaginary parts of
 a complex function are related to each other
by the Hilbert transforms.
Let
\beqn
\label{eq:PHY}
 \Phi ( \mu ) \equiv
    \frac{1}{2\pi N} \frac{\partial}{\partial \mu}
 \left( \log A + \log B  \right)  \;\;\;.
\eeqn
The Hilbert transform is given by
\beqn
\label{eq:Htr}
  \rho_{N} ( \mu )=  \Im \Phi \left( \mu  \right) = - P
   \int   \frac{ d \mu^{\prime} }{ \pi}
 \frac{  \Re  \Phi \left( \mu^{\prime}
 \right) } {  \mu^{\prime} -    \mu }
 = - \frac{1}{2\pi}  \int_{ \Gamma^{(+)} + \Gamma^{(-)} }
     d  \mu^{\prime}
 \frac{ \Re \Phi \left( \mu^{\prime}
 \right) } {  \mu^{\prime} -   \mu }  \;\;\;.
\eeqn
 Here $  \Gamma^{(+)}$  $(\Gamma^{(-)}) $   denotes
 a path  which runs along  the real axis and which avoids $\mu$
  from  above  ( below ).
  We find, for large $N$,
\beqn
  \rho_{N} (\mu) = - \frac{i \sqrt{2}}{8}
  \int_{ \Gamma^{(+)} + \Gamma^{(-)} }
 \frac{dx}{2\pi i} \frac{ \tanh \left(-\pi N (\sqrt{2}/\lambda)
g(x) \right)  g^{\prime} (x) }{ x- k_{comp}^{2} } \;\;\;,
\eeqn
 where $ k_{comp}^{2} \equiv -\mu \lambda/4 = 1- k_{w}^{2}$ and
\beqn
   g^{\prime} (x) = F( 1/2,1/2 ;1 \mid x) = \frac{2}{\pi} K \left(
  \sqrt{x} \right) = \frac{2}{\pi} \int^{\pi/2}_{0} d \theta
 \frac{1}{ \sqrt{1-x \sin^{2} \theta}  } \;\;\;.
\eeqn
  Picking up an infinite number
 of poles  along the positive and negative  imaginary axes, we
find
\beqn
\label{eq:answerN}
   \rho_{N} \left( \mu \right) &=& - \frac{ \lambda}{4\pi} \Im
  \left( \frac{1}{N} {\displaystyle \sum_{ \ell=0}^{\infty} }
\frac{1}{iy_{\ell} - k_{comp}^{2} } \right) \;\;\;, \\
  iy_{\ell} &\equiv& g^{-1} \left( i \frac{ \sqrt{2} \lambda}{2}
 \frac{(\ell +1/2)}{N} \right) \;\;\;, \;\; \ell = 0,1,2, \cdots
  \nonumber
\eeqn
 where $g^{-1}(x)$ is the inverse function of $g(x)$. For later purpose,
 we also write it as
\beqn
\label{eq:anotherN}
   \rho_{N} \left( \mu \right) &=& - \frac{\lambda}{4\pi}
  \lim_{L \rightarrow \infty} \Im
  \left(  {\displaystyle \sum_{\ell =0}^{L} }
\frac{1}{Niy_{\ell} - \tilde{k}_{comp}^{2} } \right) \;\;\;,
\eeqn
  where $ \tilde{k}_{comp}^{2} \equiv  N  k_{comp}^{2}. $
  For given finite $N$, the infinite series eq.~(\ref{eq:answerN})
 is convergent:  the function  $g^{-1}(x)$ grows  faster than $x$,
 as  $x$ goes to infinity according to eq.~(\ref{eq:ghyper}).
  (We will discuss this in more detail  in the next subsection.)
  This may be regarded as a consequence of the
 fact that the system has a stable ground state.

\subsection{  Finiteness in the scaling limit}

Let us now take the scaling limit, which
  is  $N \rightarrow \infty$   while keeping
 $ \tilde{k}_{comp}^{2} $ large but finite
 in eq.~(\ref{eq:anotherN}).   The limit is defined to send
  $L$ to infinity  first and subsequently $N$ to infinity.
\beqn
\label{eq:limit}
     \rho \left( \mu \right) &\equiv& - \frac{\lambda}{4\pi}
  \lim_{N \rightarrow \infty}   \lim_{L \rightarrow \infty} \Im
  \left(  {\displaystyle \sum_{\ell =0}^{L} }
\frac{1}{Niy_{\ell} - \tilde{k}_{comp}^{2} } \right) \;\;\;\\
          & \equiv & - \frac{\lambda}{4\pi}
  \lim_{N \rightarrow \infty}   \lim_{L \rightarrow \infty}
  q \left(N, L \right) \;\;\;.  \nonumber
\eeqn
 We  would like to prove that the  limit in eq.~(\ref{eq:limit}) is finite.
  First,  note that
\beqn
\label{eq:qnlseries}
  q(N,L) &=& - {\displaystyle \sum_{\ell=0}^{L} }
 \frac{N \Im    g^{-1} \left( i \frac{ \sqrt{2} \lambda}{2}
 \frac{(\ell +1/2)}{N} \right)  } {
  \left(  N  \Re g^{-1} \left( i \frac{ \sqrt{2} \lambda}{2}
 \frac{(\ell +1/2)}{N} \right) - \tilde{k}_{comp}^{2} \right)^{2}
  +  \left( N \Im    g^{-1} \left( i \frac{ \sqrt{2} \lambda}{2}
 \frac{(\ell +1/2)}{N} \right)  \right)^{2}   }
 \nonumber   \\
  &\equiv&  -  {\displaystyle \sum_{\ell=0}^{L} }  a_{\ell}
\left( N \right) \;\;\;.
\eeqn
 For $  N$  fixed and $\ell$ large, we know that
\beqn
\label{eq:re}
   \Re g^{-1} \left( i \frac{ \sqrt{2} \lambda}{2}
 \frac{(\ell +1/2)}{N} \right)  \sim  \alpha
   \left( \frac{(\ell +1/2)}{N} \right)^{1+ \delta} \;\;\;,
\eeqn
  where $\alpha$ is a  real constant, and  $\delta =1$ up to
 logarithmic corrections to the power
 law behavior   \footnote{
 This follows from the connection formula
  of  $ F( \alpha, \beta ; \gamma  \mid x) $ with
 $  F( \alpha^{\prime}, \beta^{\prime} ; \gamma^{\prime}
 \mid  \frac{1}{x})$, which provides the asymptotic expansion.
  The coefficient is divergent due to the logarithmic singularity.
 The only thing we need in what follows, however, is that
  $\delta$ be a positive number independent of $N$.}.
  We can set a bound on the imaginary part:
\beqn
\label{eq:im}
   \Im g^{-1} \left( i \frac{ \sqrt{2} \lambda}{2}
 \frac{(\ell +1/2)}{N} \right)    <   \beta
   \left( \frac{(\ell +1/2)}{N} \right)^{1+ \delta} \;\;\;,
\eeqn
  with $\beta$ a positive number.
   We find  therefore
\beqn
\label{eq:bound}
  \mid  a_{\ell} \left( N \right) \mid~~ < ~~
\frac{  N  \beta   \left( \frac{(\ell +1/2)}{N} \right)^{1+ \delta}  }
 {  \left( N \alpha
   \left( \frac{(\ell +1/2)}{N} \right)^{1+ \delta} - \tilde{k}_{comp}^{2}
  \right)^{2}  }
   \equiv    \tilde{ a_{\ell} } \left( N \right) \;\;\;.
\eeqn
  We use  the convergence (divergence) criterion
\beqn
\label{eq:crit}
   \lim_{\ell \rightarrow \infty} \ell \log \left( \frac{
  \tilde{a}_{\ell}  \left( N \right) }
{  \tilde{a}_{\ell+1}  \left( N \right) } \right)
    > 1 ~~\left( < 1 \right)~~~~
   \Longleftrightarrow ~~  {\rm convergent~~ (
  divergent)}  \;\;\;.
\eeqn
  After a  short calculation, we  obtain
\beqn
\label{eq:converge}
     \lim_{\ell \rightarrow \infty} \ell \log \left( \frac{
  \tilde{ a }_{\ell}  \left( N \right) }
{  \tilde{ a }_{\ell+1}  \left( N \right) } \right)
     = 1+  \delta~ >~  1   \;\;\;.
\eeqn
 The infinite series
 $  {\displaystyle \sum_{\ell =0}^{\infty} }
 \tilde{a}_{\ell} \left( N \right) $
 is convergent and  correspondingly the
   $ {\displaystyle  \lim_{L \rightarrow \infty} }
  q \left(N, L \right)$  as well as eq.~(\ref{eq:answerN})
  as we stated earlier.
   It is important to note
 that the right hand side of eq.~(\ref{eq:converge})
 is a number independent of $N$ and  therefore
\beqn
\label{eq:convergeN}
     \lim_{N \rightarrow \infty}   \lim_{\ell \rightarrow \infty}
 \ell \log \left( \frac{
  \tilde{ a }_{\ell}  \left( N \right) }
{  \tilde{ a }_{\ell+1}  \left( N \right) } \right)
     = 1+  \delta ~ > ~  1   \;\;\;.
\eeqn
 This proves the convergence  of $  {\displaystyle
  \lim_{N \rightarrow \infty} } {\displaystyle  \sum_{\ell =0}^{\infty} }
 \tilde{ a }_{\ell}  \left( N \right) $
  and hence the finiteness of
$  {\displaystyle \lim_{N \rightarrow \infty}}$
 ${\displaystyle   \lim_{L \rightarrow \infty}}$
  $q \left(N, L \right)$, which is what we wanted to show.

On the other hand,
it is easy to see that   eq.~(\ref{eq:limit})
 contains  the asymptotic series representing
  string perturbation theory. Keep $L$ fixed  while sending
 $N$ to infinity first, and subsequently send $L$ to infinity.
 The function $g^{-1}(x)$ is linearized in this limit
 and we recover the  asymptotic
 series \cite{c=1} coming from the inverted harmonic
 oscillator;
\beqn
 \label{eq:divergent}
     \rho_{as} \left(  \tilde{ \mu } \right) & \equiv& - \frac{\lambda}{4\pi}
   \lim_{L \rightarrow \infty}
  \lim_{N \rightarrow \infty}  \Im
  \left(  {\displaystyle \sum_{\ell =0}^{L} }
\frac{1}{iNy_{\ell} - \tilde{k}_{comp}^{2} } \right)
 \;\;\;  \\
 ~ &=&   \frac{1}{2 \sqrt{2} \pi}  \left[  \log (1/k^{2}_{comp}) +
  \sum_{m=1}^{\infty}
  \left( 2^{2m-1} -1 \right) \frac{ \mid B_{2m} \mid }{m}
  \frac{1}{ \tilde{ \xi }^{2m}}  \right]  \;\;\;, \label{eq:Bernoulli}
\eeqn
 where  $\tilde{ \xi } = - \frac{\sqrt{2}}{2} \tilde{\mu}$.
    The  perturbative piece  is in fact contained in the original
 expression.  Eqs.~(\ref{eq:divergent}), (\ref{eq:Bernoulli})  tell
 us that
 this can be exhibited by blowing up the region
 of the singularity in eq.~(\ref{eq:limit}), which is achieved by the
  double scaling limit.

It is instructive to compute
 $  {\displaystyle
  \lim_{N \rightarrow \infty}  }
  a_{\ell}  \left( N \right)$:
\beqn
   {\displaystyle
  \lim_{N \rightarrow \infty}  }
  a_{\ell}  \left( N \right) = \frac{  \frac{\sqrt{2}}{2} \left(
  \ell + 1/2 \right) }{ \left( \frac{\sqrt{2}}{2} \left(
  \ell + 1/2 \right)  \right)^{2} +  \left(  \tilde{k}^{2}_{comp} \right)^{2} }
   \equiv b_{\ell} \;\;\;.
\eeqn
  We find
\beqn
       \lim_{\ell \rightarrow \infty} \ell \log \left( \frac{
   b_{\ell}   }
{  b_{\ell+1} } \right)
     = 1   \;\;\;, \;\;
       \lim_{\ell \rightarrow \infty}  \left[ \ell  \left( \frac{
   b_{\ell}  }
{ b_{\ell+1}   }  -1 \right) -1 \right] \log \ell
     = 0~<~1   \;\;\;,
\eeqn
 confirming that  eq.~(\ref{eq:Bernoulli}) is  indeed divergent.

\subsection{Ground state energy in the scaling limit}

 So far, we have obtained a finite expression for the density
 of states as a function of  $\tilde{ \mu }$ in the scaling limit.
  As a model of the $1+1$ dimensional string, the physical quantities
  must be expressible as a function of the renormalized cosmological
  constant.   We  derive  below
    the ground state energy in the scaling
  limit as an implicit function of
 the renormalized cosmological constant $\tilde{x}$.

  Consider the normalization condition (eq.~(\ref{eq:norm})).
 The solution is written as  $ \mu = \mu_{N} (\lambda)$.
  The function   $\mu_{N} (\lambda)$ satisfies
 $  {\displaystyle \lim_{N \rightarrow \infty} }  \mu_{N} (\lambda_{c}) = 0.$
  In order to remove  the nonuniversal piece from  eq.~(\ref{eq:norm}),
  let us study the response of eq.~(\ref{eq:norm}) to
 the small variations of  $\lambda$  and  $\mu$  from
 their critical values    $\lambda_{c}$ and  $ \mu_{N} (\lambda_{c})$:
\beqn
  \mu_{N} (\lambda_{c})  \rightarrow   \mu_{N} (\lambda_{c})
  +   \delta \mu   \;\;\;, \;\;
 \lambda_{c} \rightarrow    \lambda_{c}  +  \delta \lambda
  \;\;\;.
\eeqn
  We find
\beqn
  0=   \rho_{N}  \left(  \mu = \mu_{N} \left(\lambda \right) ;
  \lambda_{c} \right)  \delta \mu +
 \frac{ \partial {\cal  N}_{N} }{ \partial \lambda} \mid_{ \stackrel{\mu
=  \mu_{N} \left(\lambda_{c} \right)}{ \lambda = \lambda_{c}}  }
  \delta \lambda \;\;\;.
\eeqn
   Multiply this equation by $N$ and take the scaling limit.
  We find
\beqn
\label{eq:muasx}
  \frac{ d \tilde{x}}{d \tilde{\mu}} = \rho \left( \tilde{ \mu },
 \lambda_{c} \right) \;\;\; {\rm  or}  \;\;\;
  \tilde{x}  = \int_{0}^{ \tilde{ \mu }}  d \tilde{\mu}^{\prime}
  \rho \left( \tilde{ \mu }^{\prime}, \lambda_{c} \right) \;\;\;.
\eeqn
   This determines  $\tilde{\mu}$ as a function of $ \tilde{x}$.

   As for the  ground state energy, we write it as
\beqn
    E_{N} \left( \lambda \right)-  \int^{0}_{-4/\lambda}
  d \mu^{\prime}  \mu^{\prime}  \rho_{N} ( \mu^{\prime} ;\lambda) =
      \int_{0}^{ \mu = \mu_{N} \left( \lambda \right)}
  d \mu^{\prime}  \mu^{\prime}  \rho_{N} ( \mu^{\prime} ;\lambda) \;\;\;.
\eeqn
   We find that the rescaled ground state energy is given by
\beqn
\label{eq:resgr}
 E_{gr} \left( \tilde{x} \right) \equiv
  \lim_{  \stackrel{N \rightarrow \infty}
{\lambda \rightarrow \lambda_{c}}  }
  N^{2} \left(    E_{N} \left( \lambda \right)-  \int^{0}_{-4/\lambda}
   d \mu^{\prime}  \mu^{\prime}  \rho_{N} ( \mu^{\prime} ;\lambda)
  \right)
\nonumber
\\
= \int_{0}^{ \tilde{ \mu } =  \tilde{ \mu } ( \tilde{x} ) }
 d  \tilde{ \mu }^{\prime}  \tilde{ \mu }^{\prime} \rho
(  \tilde{ \mu }^{\prime} ;\lambda_{c} )  \;\;\;.
\eeqn

  To recapitulate, the set of  equations
\beqn
\label{eq:recap}
     \rho \left( \mu \right) &=& - \frac{\lambda}{4\pi}
  \lim_{N \rightarrow \infty}   \lim_{L \rightarrow \infty} \Im
  \left(  {\displaystyle \sum_{\ell =0}^{L} }
\frac{1}{Niy_{\ell} - \tilde{k}_{comp}^{2} } \right) \;,\;
    ~~~~~ \nonumber   (\ref{eq:limit})  \\
  iy_{\ell} &=&  g^{-1} \left( i \frac{ \sqrt{2} \lambda}{2}
 \frac{(\ell +1/2)}{N} \right) \;\;\;,
  \nonumber  \\
  \tilde{x}  &=& \int_{0}^{ \tilde{ \mu }}  d \tilde{\mu}^{\prime}
  \rho \left( \tilde{ \mu }^{\prime}, \lambda_{c} \right) \;\;\;,
 \;\;\;  \Longrightarrow \;\;\;  \tilde{ \mu } =  \tilde{ \mu }
 ( \tilde{x} )\;\;,~~~~~   \nonumber (\ref{eq:muasx})     \\
 E_{gr} \left( \tilde{x} \right) &=& \int_{0}^{ \tilde{ \mu } =
  \tilde{ \mu } ( \tilde{x} ) }
 d  \tilde{ \mu }^{\prime}  \tilde{ \mu }^{\prime} \rho
(  \tilde{ \mu }^{\prime} ;\lambda_{c} )~~~~~~~   \nonumber  (\ref{eq:resgr})
    \;\;\;.
\eeqn
  determines the rescaled ground state energy unambiguously
  as a function of the renormalized cosmological constant.
    Eqs.~(\ref{eq:limit}),(\ref{eq:norm}) and
  (\ref{eq:resgr}) could be
 studied numerically to obtain finite results.
  On the other hand, the asymptotic expression for the ground
  state energy  $E_{gr}^{(as)} (\tilde{x})$ is obtained by replacing
  $\rho$ by $\rho_{as}$ in the above set of equations.
  The divergent series eq.~(\ref{eq:Bernoulli})  must be amputated
 up to the optimal term to render   $E_{gr}^{(as)} (\tilde{x})$  finite.
  The difference of the  corresponding two point functions, for example,
\beqn
  \left(  \frac{d}{dx} \right)^{2} \left( E_{gr}(\tilde{x})
  -E_{gr}^{(as)} (\tilde{x}) \right)
\eeqn
  can then be studied to determine the nonperturbative contributions
 of the susceptibility exponent.

\section{Conclusion}
We have shown that the one-plaquette model offers a
stable and well-defined arena in which to study nonperturbative effects in
$1+1$ dimensional string theory.  We compute an exact expression for
the density of states of the one-plaquette model.
  Our results appear to be the first example in a $1+1$ dimensional
  string where nonperturbative effects are given in an explicit form.
In this paper, we have focussed on the (static) ground state properties.
It would be interesting to see how the model reveals nonperturbative
effects in, for example, (time-dependent) scattering processes.

\section{ Acknowledgements}
This work is supported in part by the Robert A. Welch Foundation,
 NSF grant PHY 9009850  and Grant-in-Aid for  Scientific Research
  $(05640347)$  from the Ministry of Education, Japan.

\newpage

\appendix

\section{ }

  In this appendix, we outline the derivation of the density of states
 from the WKB wave function for the case of the inverted harmonic oscillator.
 The basic idea of the derivation is seen in \cite{Kaz}.
 Consider the eigenvalue problem
\beqn
 \hat{h}_{M}\psi &=&  \mu \psi  \;,\;\; \nonumber \\
  h_{M} &\equiv&  -\frac{1}{2N^{2}} \left( \frac{d}{dx} \right)^{2}
  + V\left( x\right) \;\;,\;\; V\left( x\right) = -   x^{2}/2 \;\;\;.
\eeqn
  subject to the boundary condition
\beqn
\label{eq:walls}
  \psi (x= \pm \Lambda) =0 ,
\eeqn
 which is accomplished by
 putting infinite walls at  $x= \pm \Lambda$.
We will be applying the semi-classical approximation and the
 expansion by $\mu^{1/2}/x$ which
are valid in the region
\beqn
   N x^{2} >> 1 \;\;\;, \;\;\;  x^{2} >> \mid \mu \mid \;\;\;.
\eeqn
Let us prepare  a set of two wave functions
  $ \psi^{  \left(\pm \right)} \left( x ; \mu
 \right)$ which are complex conjugate to
 each other. Their WKB form in the region $ \epsilon >V(x)$ is
\beqn
\label{eq:invwkb}
   \psi_{WKB}^{\left(\pm \right)} = \left( 2  \mu
 + x^{2} \right)^{-1/2}
   e^{ \pm i  N ( x^{2}/2 +   \mu \log x  ) } \;\;\;.
\eeqn
As the potential is real, the eigenfunctions can be chosen to be real:
\beqn
 \psi^{\left(R \right)}\left( x ;  \mu \right)
   &=& B \psi^{ \left(+ \right)}\left( x ;  \mu \right)
     + B^{*} \psi^{\left(- \right)}
    \left( x ;  \mu \right) \;  ,\;\;\;\;\;\;\;\;\; x > 0
    \;\;\; \\
 \psi^{\left(L \right)}\left( x ;  \mu \right)
   &=& A \psi^{ \left(+ \right)}
\left( \mid x \mid ;  \mu \right)
     + A^{*} \psi^{  \left(- \right)}
    \left( \mid x \mid ;
  \mu \right) \;  ,\;\;\; x < 0
 \eeqn
 The semiclassical expressions   for
   $  \psi^{\left(R \right)}\left( x ;  \mu \right)$
 and
 $ \psi^{\left(L \right)}\left( x ;  \mu \right)$
 are  given   by
\beqn
  \psi^{\left(R \right)}_{WKB} \left( x ;  \mu \right)
   &=& 2 \left( 2\mu + x^{2} \right)^{-1/2}
  \mid B \mid  \cos \left( \arg B +  N( x^{2} /2 + \mu
  \log x  ) \right)  \;\;\;, \nonumber \\
 \psi^{\left(L \right)}_{WKB} \left( x ;  \mu \right)
   &=& 2 \left( 2\mu + x^{2} \right)^{-1/2}
  \mid A \mid  \cos \left( \arg A +  N ( x^{2} /2 + \mu
  \log x )  \right)  \;\;\;.
\eeqn
 The boundary condition eq.~(\ref{eq:walls}) provides
 a quantization of energy levels
\beqn
\label{eq:wallmat}
  \arg A +  N \left( \frac{\Lambda^{2}}{2} + \mu \log \Lambda \right)
   &=&  \left( n_{L} + \frac{1}{2} \right) \pi \;\;\;,\;\;\;
    n_{L}  \in {\cal Z} \nonumber \\
  \arg B +  N \left( \frac{\Lambda^{2}}{2} + \mu \log \Lambda \right)
   &=&  \left( n_{R} + \frac{1}{2} \right) \pi \;\;\;, \;\;\;
     n_{R}  \in {\cal Z} \;\;\;.
\eeqn

 The matching conditon  is derived in the same way as  was done in the
 text for the
 one-plaquette model and we find
\beqn
  A =  e^{-\pi N \mu} B \;\;\;.
\eeqn
Through the normalization condition
\beqn
 1=   \int^{\Lambda} \psi^{(R)} \left( x ; \mu
 \right)^{2} +
  \int_{-\Lambda} \psi^{(L)} \left( x ; \mu
 \right)^{2}  \;\;\;,
\eeqn
  we obtain
\beqn
\label{eq:dlogA}
    \frac{1}{2\pi N}  \frac{\partial}{\partial \mu}
 \left(  \log \mid A \mid  + \log \mid B \mid  \right)
     =  -\frac{1}{2} \tanh \pi N \mu  \;\;\;.
\eeqn
   From eq.~(\ref{eq:wallmat}) the density of states  is
\beqn
  \rho_{N} \left( \mu \right) =  \frac{1}{2\pi N}
  \frac{\partial}{\partial \mu}
 \left(  \arg  A   + \arg B  \right)  +  \frac{1}{\pi}  \log \Lambda  \;\;\;,
\eeqn
 and  is found to be
\beqn
\label{eq:pres}
  \rho_{N} \left( \mu \right)
 = - \frac{1}{2\pi}  \int_{ \Gamma^{(+)} + \Gamma^{(-)} }
     d  \mu^{\prime}
 \frac{    -\frac{1}{2} \tanh \pi N \mu^{\prime}
 + \frac{1}{\pi} \log \Lambda  }
 {  \mu^{\prime} -   \mu }  \;\;\;
\eeqn
  from eq.~(\ref{eq:dlogA}) and
 the Hilbert transform.  The last integral is  divergent as well
 as  $\Lambda$-dependent however
 and should  be viewed  only as a prescription
 to reproduce the asymptotic series
  quoted in the text.


\newpage



\begin{thebibliography}{99}

\bibitem{BKDSGM}
E.~Brezin and V.~Kazakov, {\sl Phys. Lett.} {\bf 236B} (1990)
144;~M.~Douglas and S.~Shenker, {\sl Nucl. Phys.}~{\bf B335} (1990) 635
;~D.~Gross and A.~Migdal, {\sl Phys. Rev. Lett.} {\bf 64} (1990)
127; {\sl Nucl. Phys.}~{\bf B340}(1990)333.


\bibitem{c=1}
D. Gross and N. Miljkovic, {\sl Phys. Lett.}{\bf 238} (1990) 217.
E. Brezin, V. Kazakov and A. B. Zamalodchikov, {\sl Nucl. Phys.} {\bf 338}
(1990) 673.  P. Ginsparg and J. Zinn-Justin, {\sl Phys. Lett.} {\bf 240}
(1990) 333.

\bibitem{Review}
  I.~Klebanov,  lectures delivered at the ICTP Spring School on String
  Theory and Quantum Gravity, Trieste, (April 1991).

\bibitem{Kaz}
V.~Kazakov, in Cargese workshop on ``Random Surfaces and Quantum
Gravity'' (May 1990) edited by O. Alvarez, E. Marinari and P. Windey.

\bibitem{Moore}
  G.~Moore, {\sl Nucl. Phys.} {\bf B368} (1992) 557. See also, P. Ginsparg
 and G. Moore, TASI $1992$ Lecture notes, and references therein.

\bibitem{GrossWitten}
  D.~Gross and E.~Witten, {\sl Phys. Rev.} {\bf D21} (1980) 446.

\bibitem{umm}
 V.~Periwal and D.~Shevitz, {\sl Phys. Rev. Lett.} {\bf 64} (1990)
 1326; C.~Crnkovic, M.~Douglas and G.~Moore, {\sl Nucl. Phys.}
{\bf B360} (1991) 507;

\bibitem{neub}
H. Neuberger, {\sl Nucl. Phys.} {\bf B179} (1981) 253.
H. Neuberger, {\sl  Nucl. Phys.}{\bf B340} (1990) 703.

\bibitem{dem}
K. Demeterfi and C.-I. Tan, Mod. Phys. Lett. {\bf A5} (1990) 1562.

\bibitem{Wadia}
S.~Wadia, {\sl Phys. Lett.} {\bf 93B} (1980)403.

\bibitem{JS}
A.~Jevicki and B.~Sakita, {\sl Phys. Rev.} {\bf D22} (1980)467.

\bibitem{qcd2old}
A. Migdal, Zh. Eksp. Teor. Fiz. {\bf 69} (1975) 810. V. Kazakov and
I. Kostov, Nucl. Phys. {\bf B176} (1980) 199.
I. Kostov, Phys. Lett. {B138} (1984) 191.

\bibitem{qcd2new}
B. Rusakov, Mod. Phys. Lett. {\bf A5} (1990) 693. D. Gross, PUPT-1356 (1992).
J. Minahan, UVA-HET-92-10 (1992).  D. Gross and W. Taylor, PUPT-1376,
and PUPT-1382 (1993). J. Minahan and A. Polychronakos, hepth/9303153 (1993).
M. Douglas, RU-93-13 (1993). M. Douglas and V. Kazakov, LPTENS-93/20 (1993).

\end{thebibliography}
\end{document}